

\magnification=1200
\global\newcount\meqno
\def\eqn#1#2{\xdef#1{(\secsym\the\meqno)}
\global\advance\meqno by1$$#2\eqno#1$$}
%
\global\newcount\refno
\def\ref#1{\xdef#1{[\the\refno]}
\global\advance\refno by1#1}
\global\refno = 1
\vsize=7.5in
\hsize=5.6in
\tolerance 10000
%
%

\def\bx{\partial^2}

\def\half{{1\over2}}

\def\bx{\partial^2}

\def\half{{1\over2}}

\baselineskip=0.1cm

\baselineskip 12pt plus 1pt minus 1pt
\vskip 2in
\centerline{{\bf RENORMALIZATION GROUP APPROACH TO SCALAR THEORY}
\footnote{*}{This work is
supported in part by funds
provided by the U. S. Department of Energy (D.O.E.) under contract
\#DE-FG05-90ER40592.}}
\vskip 24pt
\centerline{Sen-Ben Liao and Chengqian Gong}
\vskip 12pt
\centerline{\it Department of Physics}
\centerline{\it Duke University }
\centerline{\it Durham, North Carolina\ \ 27708\ \ \ U.S.A.}
\vskip 12pt
\vskip 2.5in
\vskip 24 pt
\baselineskip 12pt plus 2pt minus 2pt
\centerline{{\bf ABSTRACT}}
\medskip
Effective potential for scalar $\lambda\phi^4$ theory is obtained using
the exact renormalization group method which includes both the usual
one-loop contribution as well as the
dominant higher loop effects. Our numerical calculation
indicates a breakdown of naive one-loop result for sufficiently large
renormalized coupling constant.

\vskip 24pt
\vfill
\noindent Duke-TH-94-63\hfill April 1994
\eject

\vskip 2in
\medskip
\nobreak
\xdef\secsym{}\global\meqno = 1
\medskip

Renormalization group (RG) has been a powerful non-perturbative
method in probing how
fundamental laws of physics are modified with varying observational
length scale \ref\wils. Starting from a bare action $S$
characterizing the system at a typical microscopical length scale
$\Lambda^{-1}$, where $\Lambda$ is
the momentum cut-off of the theory, if one wishes to examine
the physics at a scale $\tilde\Lambda < \Lambda$,
it is often desirable to consider a low energy effective action
$\tilde S$ by a systematic elimination
of the modes between $\tilde\Lambda$ and $\Lambda$. In other words,
the cut-off is lowered to $\tilde\Lambda$ at the expense of having
to use a more complicated
action $\tilde S$ which in general also contains non-local interactions.  How
$\tilde S$ connects to the original $S$ is dictated by
Wilson's functional differential RG flow equation. Therefore, a complete
solution of the flow equation would provide a complete knowledge of
the theory at any length scale.

In the present work, we derive a RG flow equation based on the concept of
blocking transformation \ref\lp, \ref\wett.
Consider for simplicity a scalar theory described by the bare lagrangian:
\eqn\lagg{ {\cal L}={1\over 2}(\partial_{\mu}\phi)^2+V(\phi).}
Instead of using the original field variable $\phi(x)$, we introduce the
coarse-grained ``blocked variable'':
\eqn\kfiss{\phi_k(x)=\int d^4y~\rho_k(x-y)\phi(y),}
via a smearing function $\rho_k(x)$, with $k^{-1}$ being the
characteristic linear dimension of the region over which
the field averaging is performed. For simplicity, the smearing function
shall be chosen as a sharp momentum regulator:
\eqn\shcr{\rho_k(x)=\int_k^{\Lambda}{d^4p\over (2\pi)^4}
{}~e^{-ipx}\longrightarrow\rho_k(p)=\Theta(k-p).}
By splitting $\phi(x)$ into the slowly-varying background $\chi(x)$,
and the fast-fluctuating modes $\xi(x)$:
\eqn\field{\phi(p)=\cases{\chi(p),&$0 \le p \le k$ \cr
\cr
\xi(p), &$k < p < \Lambda$,  \cr }}
such that $\phi_k(p)=\rho_k(p)\phi(p)=\chi(p)$,
and by demanding that the field average $\Phi$ of a given block coincides with
the slowly varying background, one then integrates out $\xi(x)$ using the
loop expansion to obtain the blocked action $\tilde S_k[\Phi]$, which
is the effective action at the energy scale $k$. In the one loop
approximation, it takes on the familiar form:
\eqn\efact{\tilde S_k[\Phi]= S[\Phi]+~\half{\rm Tr'}{\rm ln}
\bigl[-\bx+V''(\Phi)\bigr] ,}
where the prime notation stands for derivatives taken with respect to
$\phi$, and ${\rm Tr'}$
implies that the trace in momentum space is to be carried out for
$k \le p \le \Lambda$, i.e., the modes which are to be eliminated by
blocking transformation.

The blocked action in its most general form, can however be written as
\eqn\efactt{ \tilde S_k[\Phi]=\int_x\biggl({Z_k(\Phi)\over2}
(\partial_{\mu}\Phi)^2+U_k(\Phi)+O(\partial^4)\biggr),}
where $Z_k$ is the wavefunction renormalization constant. The task of
analyzing the RG flow pattern of the theory can be simplified if one
concentrates on the blocked potential $U_k(\Phi)$ which is the
derivative-independent sector of $\tilde S_k$. In this framework,
the wavefunction renormalization is set to be unity and its small correction
can be computed using the derivative expansion technique illustrated in
\ref\fraser,\ref\liao. The lowering of
the cut-off from $\Lambda$ to $\sim k$ can be achieved in a ``smooth''
manner by first dividing the volume of momentum integration into a large
number of ``thin shells'' of width $\delta k$, each containing a small
number of modes, followed by a systematic integration of each individual
shell. This results in the following
non-linear RG evolution equation at the one-loop level:
\eqn\flow{ k{{\partial U_k(\Phi)}\over {\partial k}}=-{k^4\over 16\pi^2}
{\rm ln}\Bigl({{k^2+U_k''(\Phi)}\over {k^2+U_k''(0)}}\Bigr).}
However, since each loop integral is proportional to the volume of the
thin shell, the higher loop contributions to the functional flow
pattern are suppressed in the small
$\delta k$ limit. Hence, \flow\ can be taken as
an ``exact'' RG equation \wett.
This can be contrasted with the blocked potential ${\tilde U}_k(\Phi)$
which is derived by eliminating each individual mode independently
from one another:
\eqn\flowin{ k{{\partial {\tilde U}_k(\Phi)}\over {\partial k}}
=-{k^4\over 16\pi^2}
{\rm ln}\Bigl({{k^2+V_R''(\Phi)}\over {k^2+V_R''(0)}}\Bigr),}
where $V_R(\Phi)$ is the classical potential less the counterterm sector.
Our RG equation generated by sharp
cut-off regulator method
can also be compared with that obtained in \ref\wegner, where a
smooth decrease of the cut-off from $\Lambda$ to $e^{-t}\Lambda$ for an
arbitrarily small scale factor $t$ leads to:
\eqn\floweq{\eqalign{ {{\partial S}\over \partial t} &={1\over 2t}
\int_p^{'}\Biggl\{{\rm ln}{{\bx S}\over \partial\phi(p)\partial\phi(-p)}
-{{\partial S}\over \partial\phi(p)}{{\partial S}\over \partial\phi(-p)}
\Biggl({{\bx S}\over \partial\phi(p)\partial\phi(-p)}\Biggr)^{-1}\Biggr\}\cr
&
-\int_p\phi(p)p_{\mu}\partial_{p_{\mu}}^{'}{{\partial S}\over\partial\phi(p)}
+dS+\Bigl(1-{d\over 2}-\eta\Bigr)\int_p\phi(p){{\partial S}\over
\partial\phi(p)}+ {\rm const.}, }}
where the prime notations in the integration and the derivative indicate
respectively, that $p$ lies
in the range $e^{-t}\Lambda \le p\le \Lambda$, and that the derivative
does not act on the $\delta$ function in ${\partial S}/{\partial\phi(p)}$.
Further projecting $\phi(p)$ onto $\phi(0)=\Phi$ \ref\hasen,
\floweq\ is reduced to a RG equation for the effective potential $U_t$:
\eqn\flowe{{{\partial U_t(\Phi)}\over \partial t}={p^4\over {(4\pi)^{d/2}
\Gamma({d\over 2})}}{\rm ln}\Bigl(1+{1\over p^2}{{\bx U_t(\Phi)}\over
{\partial \Phi^2}}\Bigr)+d\cdot U_t(\Phi)+\Bigl(1-{d\over 2}-\eta\Bigr)\Phi{
{\partial U_t(\Phi)}\over {\partial \Phi}},}
which is reminiscent to what we have in \flow\ if the terms generated
from rescalings are dropped. An alternative formulation of RG and proof
of renormalizability can be found in \ref\polch.

The power of \flow\ is that any operator generated in the effective
potential as the modes are being eliminated will be kept throughout \liao.
If one is
interested in the critical phenomena and the values of critical
exponents, the complicated RG flow equation can be simplified in the
vicinity of fixed points. For example, for the $\lambda\phi^4$ theory
in $d=4$ near the Gaussian infrared fixed point, it suffices to retain only
the relevant $\Phi^2$ and $\Phi^4$ terms
for determining the critical exponents to high accuracy; all other terms
are irrelevant \ref\crit. However, one must remember
that operators are always classified with respect to
a particular fixed point. If a theory has, say two fixed points, one
ultraviolet and one infrared, it is
possible to generate in the effective theory an operator which becomes relevant
in the infrared while being irrelevant in the ultraviolet \liao. As the
RG trajectory reaches the crossover regime, a new classification of operators
becomes necessary since the number of relevant operators corresponds to
the number of $unstable$ directions. Failure to provide an accurate
classification would lead to substantial deformation of the RG trajectory.
In lacking of a general classification scheme, it is desirable to retain
as many new operators as possible in the effective lagrangian.
Such feature indeed can easily be incorporated in our RG approach.

In inquiring the importance of the contributions from higher loops and
the irrelevant operators in
making up the effective potential,
we shall compare $U_{k=0}(\Phi)$ generated from \flow\
with the standard one-loop Coleman-Weinberg result \ref\col,
${\tilde U}_{k=0}(\Phi)$ which ignores the impact of these terms.
For simplicity, we consider $\lambda\phi^4$ theory where
\eqn\bap{ V(\Phi)={\mu_{\Lambda}^2\over 2}\Phi^2 +{\lambda_{\Lambda}\over 4!}
\Phi^4 .}

The main message here is to report the difference between
$U_{k=0}(\Phi)$ and ${\tilde U}_{k=0}(\Phi)$ even for small renormalized
coupling constant $\lambda_R=U_{k=0}^{(4)}(0)$, and show that the
discrepancy grows with increasing $\lambda_R$. This is directly related to
the fact that the latter utilizes independent-mode approximation, and
hence neglects the $continuous$ feedback from the modes which are
being eliminated successively. As we shall see below, the most severe
error, however, comes from its truncation at the one-loop
order, thereby ignoring the higher loop terms which turn out to be large.

Our strategy consists of the following: Suppose we are given the bare mass
parameter $\mu_{\Lambda}^2$, the bare coupling constant $\lambda_{\Lambda}$
and the cut-off $\Lambda$ as the input parameters.
This allows us to determine the shape of the initial bare potential
$V(\Phi)$. In our numerical integration, all
dimensional parameters will be scaled with respect to $\Lambda\equiv 1$.
A negative $\mu_{\Lambda}^2$ ensures the
characteristic double-welled feature for $V(\Phi)$. The one-loop potential
derived from \flowin\ takes on the form:
\eqn\upk{\eqalign{{\tilde U}_k(\Phi) &={\mu_R^2\over 2}
\Phi^2\Bigl[1-{\lambda_R\over64\pi^2}\Bigl(1+{k^2\over\mu_R^2}\Bigr)\Bigr]
+{\lambda_R\over 4!}\Phi^4\Bigl(1-{9\lambda_R\over64\pi^2}\Bigr)\cr
&+{1\over64\pi^2}\Bigl[\bigl(\mu_R^2+{\lambda_R\over 2}\Phi^2\bigr)^2-k^4\Bigr]
{\rm ln}\Bigl({{k^2+\mu_R^2+\lambda_R\Phi^2/2}\over
{k^2+\mu_R^2}}\Bigr),}}
which in the $k=0$ limit, simplifies to \col:
\eqn\cwuy{{\tilde U}_{k=0}(\Phi)={\mu_R^2\over 2}\Phi^2\bigl(1-
{\lambda_R\over 64\pi^2}\bigr)+{\lambda_R\over 4!}\Phi^4\bigl(
1-{9\lambda_R\over 64\pi^2}\bigr)+
{1\over 64\pi^2}\bigl(\mu_R^2+{\lambda_R\over 2}\Phi^2\bigr)^2
{\rm ln}\Bigl({{\mu_R^2+\lambda_R\Phi^2/2}\over \mu_R^2}\Bigr).}
The above forms are deduced with the help of
the one-loop renormalization conditions:
\eqn\aac{\cases{\eqalign{\mu_{\Lambda}^2&=\mu_R^2-{\lambda_R\over 32\pi^2}
\Bigl[\Lambda^2+\mu_R^2{\rm ln}\bigl({\mu_R^2\over\Lambda^2}\bigr)\Bigr] \cr
\lambda_{\Lambda} &=\lambda_R +{3\lambda_R^2\over 32\pi^2}\Bigl[
{\rm ln}\bigl({\Lambda^2\over \mu_R^2}\bigr)-1\Bigr].\cr}}}

On the other hand, the RG improved potential $U_k(\Phi)$ is
solved numerically. In Fig. 1, the gradual
transition from a double-welled bare potential $U_{k=\Lambda}(\Phi)
=V(\Phi)$ to $U_{k=0}(\Phi)$ which has a unique minimum at $\Phi=0$
is depicted. At large $k$, $U_k(\Phi)$ and $\tilde U_k(\Phi)$ are
relatively close to one another. However, the deviation becomes more
noticeable as $k$ is lowered, as can be seen in Fig. 2.
For comparative purpose, one may simply
examine the ratio of the renormalized mass parameters,
$R={U''_{k=0}(0)/{{\tilde U}''_{k=0}(0)}}$,
where ${\tilde U}''_{k=0}(0)=\mu_R^2$. Since the resulting
renormalized coupling
constants from either approach do not differ appreciably:
${U^{(4)}_{k=0}(0)/{{\tilde U}^{(4)}_{k=0}(0)}}=1.006$, we shall simply
denote them as $\lambda_R$. We notice that even for $\lambda_R=0.1$,
the ratio of the mass parameters is $R=4.93$.
Such discrepancy can be explained by the following
arguments: Our RG approach makes use of
the ``dressed'', effective vertex
functions at each step of integration for deducing
the next lower energy improved vertex functions. Therefore,
one would naturally expect additional contributions from higher loops as well
as higher order field operators. This method is analogous to
a resummation over
daisy and superdaisy diagrams in finite temperature theory \ref\jack.
Fig. 3 shows that the accumulation of higher order field
operators only gives a small correction to $U_{k=0}(\Phi)$, thereby making
it evident that the discrepancy is largely due to
higher loop effects. One is then lead to
the inevitable conclusion that the one-loop
independent-mode approximation must break down. That is, it is insufficient to
use the one-loop $\tilde U_{k=0}(\Phi)$ as the
effective potential in the infrared regime.

Can we reconcile the perturbative result \cwuy\ with that obtained through the
``exact'' RG flow equation? Fortunately, there is one parameter which
can be tuned: the renormalized coupling constant $\lambda_R$. One sees
that the higher loop effects included in the RG approach
are all multiplied by some
power of $\lambda_R$, and only by judicious choice of very small $\lambda_R$
can their effects be safely neglected. In Fig. 4, we see that as $\lambda_R$
is decreased, the agreement between
$U_{k=0}(\Phi)$ and ${\tilde U}_{k=0}(\Phi)$ becomes better. At
$\lambda_R=0.01$, the two results differ only by $6\%$.
Improvement of the one-loop result perhaps can best be seen from
Fig. 5 in which $R\to 1$ as $\lambda_R$ becomes vanishingly small.
One therefore
concludes that the naive one-loop result ${\tilde U}_{k=0}(\Phi)$ that
ignores the impacts of higher loops can be valid
for very small $\lambda_R$. For $\lambda_R$ not too small, one must
take into account their effects.

If one takes the cut-off $\Lambda$ seriously as part of the effective
theory, then by choosing a large yet finite $\Lambda$,
an interacting theory consistent with perturbation expansion may
be defined without confronting the complication of ``triviality''.
Nevertheless, the value $\lambda_R$ takes should
be checked by our improved RG method to ensure the reliability of
one-loop perturbative result. For a given $\Lambda$, we shall denote
by $\lambda_0$
the coupling constant which results in a $20\%$ difference between our RG
method and the standard one-loop integration, i.e.,
$R=1.2$ for a given $\lambda_0$. A $20\%$
difference, in our opinion, still lies within the limit of tolerance for
perturbation. In Fig. 6, the relation between $\lambda_0$ and cut-off
$\Lambda$ is illustrated in logarithmic scale. One easily sees that
the larger the $\Lambda$, the smaller $\lambda_0$ must be used
in order to trust the simple one-loop calculation.

We also comment on the sensitivity of $R$ to the choice of cut-off
$\Lambda$. Eq. \aac\ shows that the finite renormalized mass parameter
$\mu_R^2$ results from cancellation of two large numbers, namely, the
$\Lambda^2$-dependent counterterm and the bare mass parameter
$\mu_{\Lambda}^2$. Therefore, even a small adjustment of $\Lambda$
can lead to a substantial change in $\mu_R^2$.
By integrating \flow\ with slight variation of
initial choice of $\Lambda$, we find linear dependence of mass ratio $R$
on $\Lambda$. As illustrated in Fig. 7, the
slope of the line can be approximated by
${\lambda_R\Lambda\over {16\pi^2\mu_R}}$, which agrees with \aac.
On the other hand, we also see from Fig. 7 that the mass parameter
obtained using the RG
flow equation \flow\ with $\Lambda=0.994$ is the same as the simple one-loop
result with $\Lambda=1$. Interpreting the result differently, we say
that the higher loop contributions can be compensated by using
a sightly higher cut-off.
This observation has yet one other implication: If $\mu_R^2$, the mass of
the scalar particle can be measured precisely, then it becomes
imperative to know the cut-off of the theory to a very high
accuracy so that when RG is applied to the microscopic lagrangian
at the cut-off level, one eventually arrives at a macroscopic lagrangian
describing a large-distance physics that agrees with experiments.
It would be interesting to explore the dependence of the Higgs field mass
in the Standard Model on the choice of $\Lambda$ \ref\hasenfr.
Such peculiar sensitivity is only characteristic of the scalar theory,
and should not be expected in gauge theories where only
logarithmic divergences appear.

In summary, we have introduced in this paper an improved RG flow
equation whose non-perturbative nature takes into consideration the additional
dominant loop effects. It is concluded that one-loop result is inadequate
unless $\lambda_R$ is set to be small or
higher loop effects are included. Typically
one chooses $\lambda_R < 0.02$ for $\Lambda^2/\mu_R^2=10^6$
to safely ignore higher loops.
Our conceptually simple yet powerful non-perturbative method
is now being implemented to systems at finite temperature \ref\finite.
With our approach, daisy, superdaisy and higher order effects are
automatically included. In addition, upon employing a suitable choice of
smearing function in the proper-time formalism \ref\gauge,
the RG evolution equations for gauge theories can also be
generated while preserving gauge symmetry.

\medskip
\bigskip
\centerline{\bf ACKNOWLEDGEMENT}
\medskip
\medskip
\nobreak
We thank Professor B. M\"uller critical for reading of the manuscript.
S.-B. L is grateful to Professor J. Polonyi for stimulating discussions.
\bigskip
\centerline{\bf REFERENCES}
\medskip
\medskip
\nobreak
\hang\par\noindent{\wils} K. Wilson and J.
Kogut, {\it Phys. Rep.} {\bf 12C} (1975) 75.
\medskip
\hang\par\noindent{\lp} S.-B. Liao and J. Polonyi, {\it Ann. Phys.}
{\bf 222} (1993) 122.
\medskip
\hang\par\noindent{\wett} C. Wetterich,
{\it Nucl. Phys.} {\bf B352} (1991) 529.
\medskip
\hang\par\noindent{\fraser} C. M. Fraser, {\it Z. Phys.} {\bf C28} (1985) 101.
\medskip
\hang\par\noindent{\liao} S.-B. Liao and J. Polonyi, Duke-TH-94-64, LPT 94-3.
\medskip
\hang\par\noindent{\wegner} F.J. Wegner and A. Houghton, {\it Phys. Rev}
{\bf A8} (1972) 401.
\medskip
\hang\par\noindent{\hasen} A. Hasenfratz and P. Hasenfratz,
{\it Nucl. Phys.} {\bf B270} (1986) 687.
\medskip
\hang\par\noindent{\polch} J. Polchinsky, {\it Nucl. Phys}
{\bf B231} (1984) 269.
\medskip
\par\hang\noindent{\crit} P. E. Haagensen, Y. Kubyshin, J. I. Latorre and
E. Moreno, UB-ECM-PF\#93-20;
J. Zinn-Justin, {\it Quantum Field Theory and Critical Phenomena},
Oxford University Press (1989).
\medskip
\hang\par\noindent{\col} S. Coleman and E. Weinberg,
{\it Phys. Rev.} {\bf D7} (1973) 1888.
\medskip
\par\hang\noindent{\jack} L. Dolan and R. Jackiw, {\it Phys. Rev.}
{\bf D9} (1974) 3320.
\medskip
\hang\par\noindent{\hasenfr} P. Hasenfratz and J. Nager,
{\it Z. Phys.} {\bf C37} (1988) 477.
\medskip
\hang\par\noindent{\finite} S.-B. Liao and J. Polonyi, MIT-CTP-2143.
\medskip
\par\hang\noindent{\gauge} M. Oleszczuk, ``A symmetries Preserving Cut-Off
Regularization''; S.-B. Liao and J. Polonyi, Duke-TH-94-65.
\medskip
\bigskip
\centerline{\bf FIGURE CAPTIONS}
\medskip
\par\hang\noindent Fig. 1 Potentials as function of $\Phi$ for $\Lambda=10$
and $\mu_R=10^{-2}$. The solid, dotted,
and square dotted lines represent, respectively, $V(\Phi)$,
${\tilde U}_{k=0}(\Phi)$ and $U_{k=0}(\Phi)$.
\medskip
\par\hang\noindent Fig. 2 Comparisons of ${\tilde U}_k(\Phi)$
and $U_k(\Phi)$ for various values of ${k\over{\Lambda}}$, with
$\Lambda=10$ and $\mu_R=10^{-2}$.
\medskip
\par\hang\noindent Fig. 3 Contribution from higher order
field operators (solid line). Dotted line represents $U_{k=0}(\Phi)$.
\medskip
\par\hang\noindent Fig. 4 Comparisons of ${\tilde U}_{k=0}(\Phi)$
and $U_{k=0}(\Phi)$ for various values of $\lambda_R$, with
$\Lambda=10$ and $\mu_R=10^{-2}$.
\medskip
\par\hang\noindent Fig. 5 Ratio between the two mass curvatures as function
of renormalized coupling constant $\lambda_R$. Note that $R\to 1$ for
$\lambda_R\to 0$.
\medskip
\par\hang\noindent Fig. 6 Cut-off dependence of $\lambda_0$, the coupling
constant which yields $R=1.2$.
\medskip
\par\hang\noindent Fig. 7 Dependence of $R$ on the cut-off $\Lambda$.
\medskip

\par
\vfill
\end